\documentstyle[epsf,twocolumn,prl,aps]{revtex}
\begin{document}
\draft
\title{Proximity to a Nearly Superconducting Quantum Critical Liquid}
\author{Qiang-Hua Wang$^{1,2}$, and A. V. Balatsky$^{3}$}
\address{$^{1}$Department of Physics and National Laboratory of Solid State
Microstructures,\\
Institute for Solid State Physics, Nanjing University, Nanjing 210093, China}
\address{$^{2}$ Department of Physics, University of California at Berkeley,
Berkeley, CA 94720, USA}
\address{$^{3}$T-Div and MST-Div, Los Alamos National Laboratory, Los Alamos, NM\\
87545, USA}

\twocolumn[\hsize\textwidth\columnwidth\hsize\csname@twocolumnfalse\endcsname
 
\date{\today}
\maketitle
 
\begin{abstract}
The coupling between superconductors and 
a quantum critical liquid that is nearly 
superconducting provides natural interpretation for the Josephson
effect over unexpectedly long junctions, and the remarkable
stripe-spacing dependence of the critical temperature in LSCO and YBCO 
superconductors.
\end{abstract}
 

] \vskip2pc \narrowtext

In the resonating-valence-bond (RVB) theory,\cite{RVB} the superconductivity
in underdoped cuprates is thermodynamically close to a quantum spin liquid.
Although at zero doping there is a N\'{e}el ordered state, it is believed
that after sufficient doping the mobilized holes disorder the N\'{e}el state
to a quantum spin liquid. This picture is in general agreement with the
experimental data measured in underdoped samples. For example, a
quasi-particle pseudo-gap forms much above the superconducting transition
temperature $T_{c}$, and both the pseudo-gap and the energy gap below $T_{c}$
exhibit the $d_{x^{2}-y^{2}}$-wave symmetry in the momentum space.\cite
{pseudogap} Indeed, in a mean field solution to the RVB theory,\cite{Kotliar}
the pseudo-gap follows from preformed spin pairs, and the condensation of
these pairs, which is determined by the hole density $x$, defines the
superconducting transition. The mean field solution captures the major
feature of the underdoped superconductors, but it was accepted with
skepticism because of the neglected gauge fluctuations. \cite{gauge}
However, it was recently realized that the gauge fluctuations can be
integrated out rigorously in the long wave length and low energy limit and a
renormalized theory can be obtained.\cite{DHLee} In this respect, the notion
of a quantum spin liquid is pertinent in the underdoped regime.

In the cuprates, superconductivity is developed beyond a critical doping $%
x_{c}(>x_{N})$, and the transition temperature $T_{c}$ increases with $x$ at 
$x>x_{c}$ (but $x<x_{p}$ with $x_{p}$ the optimum doping). Since the
transition involves condensation of spin pairs, it is reasonable to believe
that the spin liquid at $x_{c}$ is a quantum critical liquid (QCL) that is
nearly superconducting.\cite{SO5} The transition to a hole-uniform
superconductor is, however, hidden by the phase separation instability at
higher $x$, where the system is likely in a mixed phase composing of
hole-rich stripes and the hole-poor critical liquid. \cite{stripe} Below we
argue that proximity to the postulated QCL is consistent with two seemingly
unrelated sets of data: the Josephson effect in unusually lengthy junctions
made of underdoped cuprates at $x$ slightly lower than $x_{c}$,\cite{Decca}
and the stripe-spacing dependence of the transition temperature of
underdoped superconductors.\cite{Yamada,Balatsky,BalatskyShen}

We first consider the general behavior of the QCL. It is known that the
Coulomb interaction is strong and long-ranged in cuprates, which frustrates
the phase separation of holes to form stripes.\cite{Emery} Therefore, we
believe that the dynamical exponent of the QCL should be $z=1$.\cite{z=1} An
immediate consequence of this exponent is that the inverse of a length has
the dimension of energy. On the other hand, since the spin pairs are well
formed, the amplitude fluctuations of the pairs are massive and can be
integrated out in principle. The important dynamics comes from the phase $%
\theta $ degrees of freedom of the pairs. Because of the isotropy in space
and (imaginary) time, a generalized Josephson scaling relation for the phase
stiffness can be obtained as follows. The Green's function for $\theta $ is,
in general, 
\begin{equation}
G(q)=q^{-(2-\eta )}Y(q\xi ),
\end{equation}
where $q=({\bf k},\omega )$ is a three-vector (${\bf k}$ is the spatial wave
vector and $\omega $ is the Matsubara frequency), $\eta $ is the anomalous
dimension, $\xi $ is the correlation length, and finally $Y(\zeta )$ is a
scaling function of $\zeta $. The phase stiffness $\rho $ (including the
superfluid density and the compressibility) is determined by \cite{book} 
\begin{equation}
\lim_{q\rightarrow 0}G(q)=b^{2}/\rho q^{2},
\end{equation}
where $b=|\langle \sqrt{x}e^{i\theta }\rangle |\sim \xi ^{-\beta /\nu }$ is
the order parameter with $\beta $ and $\nu $ being the scaling exponent for
the order parameter and correlation length, respectively. Here $x$ is the
hole density in the QCL. Thus 
\begin{equation}
\rho =b^{2}\lim_{q\rightarrow 0}q^{-\eta }Y^{-1}(q\xi )\sim \xi ^{\eta
-2\beta /\nu }=\xi ^{-(d+z-2)},
\end{equation}
where in the last step we have used the generalized hyperscaling relation $%
2\beta =\nu (d+z-2+\eta )$ with $d=2$ being the spatial dimension in our
case. Thus the desired Josephson scaling relation reads $\rho \sim \xi ^{-1}$%
. This result could also be obtained by a simple power counting as follows.
Near criticality, one expects that 
\begin{equation}
\int_{\xi ^{d+z}}\rho (\partial _{\mu }\theta )^{2}\sim 1
\end{equation}
where the integration is over a volume of $\xi ^{d+z}$ due to the isotropy
in space and time. The ansatz $\partial _{\mu }\theta \sim 1/\xi $ gives
immediately the Josephson scaling. The above analysis implies that both the
superfluid density and the compressibility scale as energy and/or inverse
length. This statement is important to discuss the finite-size scaling in
the following. With the general behavior of the QCL in mind, let us now
discuss the relevance of the QCL in the cuprates.

{\it (I) Photodoping experiment:} In a very recent and elegant experiment, 
\cite{Decca} superconducting wires are generated by photo-doping a film with 
$x$ slightly lower than $x_{c}$. The unluminated region serves as a
junction. The critical current $I_{c}$ is measured as a function of the
junction length $s$ (in the direction of current flow). It turns out that a
relation 
\begin{equation}
I_{c}\propto 1/s  \label{Eq:Ics}
\end{equation}
holds from $s=45nm$ up to $s=100nm$, a length much larger than the coherence
length $\xi _{s}$ ($\sim 1-5nm$) in the superconducting leads. This
phenomenon is difficult to understood if the junction is a normal insulator
or metal, as one would expect $I_{c}s\propto e^{-s/l}$ where $l=\xi _{s}$
for a normal insulator and $l=v_{F}/T$ for a ballistic metal. According to
the experimental setup, $l\ll 100nm$ for both cases.\cite{Decca} Therefore,
the junction may be in a special state of matter, which is arguably a QCL.
We note that the geometry of the experimental setup is two one-dimensional
(1D) superconducting wires immersed in a QCL. This is an important aspect of
the experiment. The Cooper pairs can tunnel through an extended area
connecting the two wires. Because the junction length is the only relevant
length scale, the Josephson energy scales as 
\begin{equation}
E_{J}=F(s/\xi )/s
\end{equation}
with a scaling function $F(s/\xi )$ if the junction is not exactly at
criticality. It is expected that $F(s/\xi )$ remains a constant at $s\ll \xi 
$. The above result can also be understood as follows. Since $s$ is the only
length scale, the superfluid density $\rho _{s}$ in the junction sacles as $%
1/s$ and the transverse extension of the supercurrent flow is $X_{s}\sim s$,
up to corrections from a scaling function. The Josephson energy scale is
thus given by 
\begin{equation}
E_{J}\sim \int_{sX_{s}}dxdy\rho _{s}(\nabla \theta )^{2}\sim F(s/\xi )/s,
\end{equation}
where the integration is over the area $sX_{s}$ where supercurrent flows. We
have used $\nabla \theta \sim 1/s$ for a rough estimate, and have included
correction from the finiteness of $s/\xi $ by the scaling function. Since $%
E_{J}$ determines the critical current $I_{c}$, in suitable units we have 
\begin{equation}
I_{c}s=F(s/\xi ).
\end{equation}
At $s<\xi $ we expect that $F$ is roughly a constant, so that Eq.(\ref
{Eq:Ics}) is recovered. The plot of $I_{c}s$ is nothing but a scaling
function. From the data $I_{c}s$ drops to zero quickly at $s>100nm$.\cite
{Decca} This suggests that $\xi \sim 100nm$. The decrease of $I_{c}s$ may be
related to the thermal activation of vortices in the junction. More efforts
should be devoted to explain the sharp drop of the scaling function, which
is beyond the scope of this scaling argument. On the other hand, because $%
z=1 $, we expect that $\xi \sim 1/T$ in the critical regime, so that the
scaling relation can be rewritten as 
\begin{equation}
I_{c}s=F(Ts).
\end{equation}
An important prediction follows from this function: Instead of varying $s$,
one could vary the temperature $T$ to find the scaling behavior, provided
that the superconductivity within the wires is hardly changed. This
situation is perhaps not feasible by photodoping, but is plausible by other
techniques.

To appreciate the importance of the special geometry of the junction we
addressed so far, consider another situation. Suppose the superconducting
leads are wide enough so that effectively the QCL only lives within the
junction. Neglecting boundary corrections the Josephson energy should be
extensive in the junction width $L$. The only other length scale is $s$. By
dimension analysis, we obtain 
\begin{equation}
E_{J}\sim \int_{0}^{L}dy\int_{0}^{s}dx\rho _{s}(\nabla \theta )^{2}\sim (L/s)%
\tilde{F}(s/\xi ,s/L)/s.
\end{equation}
In the last expression, the first factor $L/s$ is extensive in $L$ but is
dimensionless. The remaining part is the characteristic energy. Here $\tilde{%
F}(s/\xi ,s/L)$ is another scaling function that is finite at vanishing
arguments. In this geometry, one would expect that 
\begin{equation}
I_{c}s=(L/s)\tilde{F}(s/\xi ,s/L).  \label{Ic2}
\end{equation}
At $s\ll \xi $ and $s\ll L$, one would expect that $I_{c}\propto L/s^{2}$,
which should be contrasted with Eq.(\ref{Eq:Ics}). This scaling form leads
us to propose the experimental test of the idea of QCL in the junction. If
one can create a wide junction by photodoping the wide stripes of material
then the critical current should follow Eq.(\ref{Ic2}). The data of Decca 
{\it et al}\cite{Decca} begin with $s>L$, and are better described by the
first geometry. (In fact, in their case, the width $L$ can not be well
defined as the wires become rounded at the tips).

The QCL should also manifest itself in the magnetic field dependence of the
Josephson critical current. Since the proliferation of vortices, each
carrying a flux quantum, would wash out superconductivity, we expect that
the magnetic field $H$ defines another length scale through $1/\sqrt{H}$.
Thus in the presence of a weak magnetic field, we expect that in a given
geometry of the junction the field dependence of the critical current should
be qualitatively given by 
\[
I_{c}(H)/I_{c}(0)=1-a(\xi \sqrt{H})^{n} 
\]
where $n$ is a universal positive exponent and $a$ a non-universal positive
constant. In the critical regime where $\xi >1/T$ the above expression
should be replaced by $I_{c}(H)/I_{c}(0)=1-a(\sqrt{H/T^{2}})^{n}$. From the
experiment of Decca et al, it turns out that $n=4$. In order to see whether
their results support the QCL scenario, it is desirable to vary both $H$ and 
$T$ and to see whether $I_{c}(H)/I_{c}(0)$ is a scaling function of $H/T^{2}$
(in the critical regime).

{\it (II) Stripes in underdoped cuprates:} Let us now discuss the relevance
of QCL in underdoped cuprates with hole-rich stripes. It was found that in
LSCO materials there exists a striking linear relation between the
superconducting critical temperature $T_c$ the incommensurate width $\delta$
near the peak at $(\pi,\pi)$ in the neutron scattering spectrum.\cite{Yamada}
Recently, a similar relation was discovered from the inelastic neutron
scattering dada in YBCO samples.\cite{Balatsky} More interestingly, it was
further argued that $T_c\propto \delta$ translates to 
\begin{equation}
T_c=hv^*/s,  \label{Eq:Yamada}
\end{equation}
where $h=2\pi\hbar$ is the Planck constant, $s$ the spacing between stripes,
and $v^*$ a material-dependent constant with the dimension of velocity.\cite
{Balatsky} Two aspects of Eq.(\ref{Eq:Yamada}) are quite unusual. First, for
a given material, $v^*$ should be constant and small ($v^*\ll v_F$ with $v_F$
being the Fermi velocity). And second, the robustness of the scaling up to
the optimum doping, where the stripes, if they existed at all, would be
densely spaced (with $s\sim 1nm$). It was conjectured that $v^*$ may be
related to an unknown collective-mode of the stripes\cite{Balatsky} and/or
the motion of heavy quasiparticles near the flat band around $(\pi,0)$.\cite
{BalatskyShen} Given the experimental results, this conjecture is an
important clue to the underlying mechanism. Here we argue that Eq.(\ref
{Eq:Yamada}) follows from dynamic stripes coupled by QCL.

We note that the stripes are metalic. They behaves as anti-phase domain
walls for the remaining anti-ferro-magnetic background.\cite{antiphase}
(Recent mean-field calculations for the t-J model indicated that in-phase
stripes are slightly more favorable energetically. However, it was believed
that quantum fluctuations might stablize anti-phase stripes.\cite{Han}) The
holes in the stripes are not strickly localized in the transverse direction.
Rather, because of quantum fluctuations, the stripes are dynamic unless
pinned by inhomogeneities. Such transverse fluctuations were believed to
develop strong superconducting correlations along the longitudinal direction
of the stripes. \cite{Krotov} Also because of the transverse fluctuations of
the stripes, the intervening region between the stripes may be driven to a
quantum critical liquid, which is almost superconducting. Combining these
considerations, we model the whole system as a superposition of
superconducting stripes and quantum critical liquids plus coupling between
them.\cite{Neto} This phenomenological model should be understood as an
effective theory in the low energy limit. It does not answer why there are
stripes and a QCL, which are the issue of a higher energy scale. Without
loss of generality, let us assume that an isolated stripe $\alpha $ can be
described by an action 
\[
S_{\alpha }=\frac{1}{2}\int dyd\tau \left[ u(\partial _{y}\theta _{\alpha
})^{2}+v(\partial _{\tau }\theta _{\alpha })^{2}\right] ,
\]
where $u$ and $v$ are the bare stiffness components with respect to space $y$
and time $\tau $, and are assumed to be independent of the the stripe
spacing $s$. (Because of transverse fluctuations of the stripes, the spacing
is only defined in an average sense.) As we argued, the stripes are coupled
by the QCL. We imagine to find the effective coupling between the stripes
mediated by the QCL. This is virtually a real-space renormalization. In the
long wave length and low energy limit, we need only to retain lowest order
couplings. Clearly, there may be corrections to the bare stiffness
components, which we believe is relatively small and is neglected in the
following. Among the other effects, the Josephson-like coupling between
nearest stripes $\alpha $ and $\beta $, 
\[
\frac{1}{2}\int dyd\tau j(\theta _{\alpha }-\theta _{\beta })^{2}
\]
is the most relevant. The coefficient $j$ can be obtained by the same
scaling arguments as that for the wide-junction geometry described above, 
\[
j=\tilde{F}(s/\xi ,s/L)/s^{2},
\]
where $L$ now denotes the characteristic length of the stripes. Again in the
long wave length and low energy limit, we can approximate $(\theta _{\alpha
}-\theta _{\beta })^{2}/s^{2}$ as $(\partial _{x}\theta )^{2}$, so that the
total action for the stripes can be put back into a continuum form in the $%
r\equiv (x,y,\tau )$ space as, 
\[
S=\frac{1}{2}\int d^{3}r\rho _{\mu }(\partial _{\mu }\theta )^{2},
\]
where $\mu $ denotes the three components, $\rho _{x}=\tilde{F}(s/\xi ,s/L)/s
$, $\rho _{y}=u/s$ and $\rho _{\tau }=v/s$. This is an anisotropic 3D
XY-model. To leading order all of the stiffness components in the new action
are inversely proportional to $s$. If the system is macroscopically
isotropic due to stripes in both directions, the global stiffness $\rho $ is
given by the geometrical mean of the three components,\cite{rho} so that 
\begin{equation}
\rho =H(s/\xi ,s/L)/s.
\end{equation}
Here $H=\tilde{F}^{1/3}$ is a scaling function given by $\tilde{F}$. The 3D
XY-model develops long range ordering as long as $\rho >0$ in thermodynamic
limit (i.e., in an infinite plane and at zero temperature). \cite{book} This
is again a quantum critical point with $z=1$ due to isotropy in space and
time. Since $\rho $ has the dimension of energy, it naturally defines the
scale of the superconducting gap. Thus at $T>\rho $ the system is in a
quantum critical state, while at $T<\rho $ it is a renormalized
superconductor. As a result, the transiton temperature is given by $%
T_{c}=\rho $, which was also indicated earlier in some stripe scenarios of
high temperature superconductivity.\cite{Tc-rho} In combination with the
behavior of $\rho $ we obtain 
\begin{equation}
T_{c}s=H(s/\xi ,s/L).
\end{equation}
For long stripes, we can simplify the expression by the approximation $%
T_{c}s=H(s/\xi ,0)$. This explains why $T_{c}$ should be determined by $s$.
The scaling function takes into account the correction due to the finiteness
of $s/\xi $. The above results follow essentially from a finite-size scaling
with respect to a QCL. It is not clear {\it a priori} that such a scaling
should work at all at the more or less microscopic stripe spacing scales.
Furthermore, the above relation does not guarantee that $T_{c}s$ will be a
constant if $H$ varies significantly at moderate arguments. However, $\xi
\sim 1/T_{c}$ at $T=T_{c}$ in the critical regime of the QCL. Consequently, 
\begin{equation}
T_{c}s=H(T_{c}s,0).
\end{equation}
This is a remarkable result, indicating that the plot of $H(T_{c}s,0)$
against $T_{c}s$ only accesses a unique point of the scaling function at $%
T_{c}s=hv^{\ast }$, with $hv^{\ast }$ determined by the equation 
\begin{equation}
hv^{\ast }=H(hv^{\ast },0).
\end{equation}
We believe that this explains the robustness of Eq.(\ref{Eq:Yamada}), even
though the scaling function $H(\zeta ,0)$ may vary with $\zeta $ due to the
finite size effect. Clearly the notion of a QCL plays an important role in
our arguments.

In summary, proximity to a quantum critical liquid that is nearly
superconducting provides appealing interpretation of the abnormal Josephson
effect and the stripe spacing dependence of $T_{c}$ in underdoped cuprates.
We also predicted the Josephson effect in a geometry different from that in
Ref.\cite{Decca}. The field dependence of the Josephson current is also
discussed. Naturally one would expect other properties of superconductors in
proximity to a quantum critical liquid. These may include scaling behavior
of the various stiffness as a function of temperature, probing frequency and
length scale.

\acknowledgments{We are grateful to Dung-Hai Lee for his insightful comments in 
the initial stage of this work, and to  I. Martin
for useful discussions. QHW was supported by the National Natural Science 
Fundation of China, the Ministry of Science and Technology of China
(NKBRSF-G19990646), and in part by the Berkeley Scholars 
Program financed by the Hutchison-Whampoa Company, Hong Kong.
Work at Los Alamos was supported by US DOE.}


\begin{references}
\bibitem{RVB}  P. W. Anderson, Science {\bf 235}, 1196 (1987).

\bibitem{pseudogap}  For a review, see, {\it e.g.}, M. Randeria,
cond-mat/9710223, and references therein.

\bibitem{Kotliar}  G. Kotliar, and J. Liu, Phys. Rev. B {\bf 38}, 5142
(1988).

\bibitem{gauge}  M. U. Ubbens and P. A. Lee, Phys. Rev. B {\bf 49}, 6853
(1998);

\bibitem{DHLee}  Dung-Hai Lee, Phys. Rev. Lett. {\bf 84}, 2694 (2000).

\bibitem{SO5}  S. C. Zhang, Science {\bf 275}, 1089 (1997).

\bibitem{stripe}  For a recent review of stripes, see, {\it e.g.}, V. J.
Emery, {\it et al}, Proc. Natl. Acad. Sci. USA {\bf 96}, 8814 (1999).

\bibitem{Decca}  R. S. Decca, H. D. Drew, E. Osquiguil, B. Maiorov, and J.
Guimpel, cond-mat/0003213.

\bibitem{Yamada}  K. Yamada, {\it et al}, Phys. Rev. B {\bf 57}, 6165 (1998).

\bibitem{Balatsky}  A. V. Balatsky, and P. Bourges, Phys. Rev. Lett. {\bf 82}
5337 (1999)

\bibitem{BalatskyShen}  A.V. Balatsky and Z.X. Shen, Science {\bf 284} 1137
(1999).

\bibitem{Emery}  V. J. Emery, {\it et al}, Phys. Rev. Lett. {\bf 64}, 475
(1990).

\bibitem{z=1}  M. P. A. Fisher, and G. Grinstein, Phys. Rev. Lett. {\bf 60},
208 (1990).

\bibitem{book}  P. M. Chaikin, and T. C. Lubensky, {\it Principles of
Condensed Matter Physics} (Cambridge University Press 1995).

\bibitem{antiphase}  L. P. Pryadko, {\it et al}, Phys. Rev. B {\bf 60} 7541
(1999), and references therein

\bibitem{Han}  Jun-Hoon Han, Qiang-Hua Wang, and Dung-Hai Lee,
cond-mat/0006046.

\bibitem{Krotov}  Y. A. Krotov, D.-H. Lee, and A. V. Balatsky,
cond-mat/9705031; Phys. Rev. B {\bf 56}, 8367 (1997).

\bibitem{Neto}  Similar starting points were invoked in, e.g., N. asselmann,
A. H. Castro Neto, and C. Morais Smith, cond-mat/005486; A. H. Castro Neto,
cond-mat/0007434.

\bibitem{rho}  This is easily seen by a duality transform.

\bibitem{Tc-rho}  V. J. Emery, S. A. Kivelson, and O. Zachar, Phys. Rev. B 
{\bf 56} 6120 (1997); J. Eroles, G.Ortiz, A. V. Balatsky, and A. R. Bishop,
Europhys. Lett. {\bf 50}, 540 (2000).
\end{references}
\end{document}